\documentclass[12pt,preprint]{aastex}

\usepackage{graphicx}

\shorttitle{Active Region Loops Observed with EIS}
\shortauthors{Warren et al.}

\begin{document}

%% ----------------------------------------------------------------------
%% --- TITLE PAGE -------------------------------------------------------
%% ----------------------------------------------------------------------

\title{Observations of Active Region Loops with the EUV Imaging
  Spectrometer on Hinode}

\author{Harry P. Warren\altaffilmark{1}, 
  Ignacio Ugarte-Urra\altaffilmark{1,2}, 
  George A. Doschek\altaffilmark{1}, 
  David  H. Brooks\altaffilmark{1,2},
  and David R. Williams\altaffilmark{3}} 

\altaffiltext{1}{Space Science Division, Naval Research Laboratory,
Washington, DC 20375}
\altaffiltext{2}{College of Science, George Mason University, 4400
University Drive, Fairfax, VA 22030}
\altaffiltext{3}{Mullard Space Science Laboratory, University College
London, Holmbury St Mary, Dorking, Surrey, RH5 6NT}

%% ----------------------------------------------------------------------
%% --- ABSTRACT ---------------------------------------------------------
%% ----------------------------------------------------------------------

\begin{abstract}
  Previous solar observations have shown that coronal loops near 1\,MK
  are difficult to reconcile with simple heating models. These loops
  have lifetimes that are long relative to a radiative cooling time,
  suggesting quasi-steady heating. The electron densities in these
  loops, however, are too high to be consistent with thermodynamic
  equilibrium. Models proposed to explain these properties generally
  rely on the existence of smaller scale filaments within the loop
  that are in various stages of heating and cooling. Such a framework
  implies that there should be a distribution of temperatures within a
  coronal loop.  In this paper we analyze new observations from the
  EUV Imaging Spectrometer (EIS) on \textit{Hinode}. EIS is capable of
  observing active regions over a wide range of temperatures
  (\ion{Fe}{8}--\ion{Fe}{17}) at relatively high spatial resolution
  (1\arcsec).  We find that most isolated coronal loops that are
  bright in \ion{Fe}{12} generally have very narrow temperature
  distributions ($\sigma_T \lesssim 3\times10^5$\,K), but are not
  isothermal. We also derive volumetric filling factors in these loops
  of approximately 10\%. Both results lend support to the filament
  models. 
\end{abstract}


%% ----------------------------------------------------------------------
%% --- INTRODUCTION -----------------------------------------------------
%% ----------------------------------------------------------------------

\section{Introduction}

High spatial resolution solar observations have shown that coronal
loops with temperatures near 1\,MK have properties that are difficult
to reconcile with physical models. Loops at these temperatures persist
for much longer than a radiative cooling time, suggesting quasi-steady
heating. The densities inferred from the observations, however, are
much higher than can be reproduced by steady, uniform heating
models. The temperature gradients along the loops are also much
smaller than predicted by the simple models.  (e.g.,
\citealt{lenz1999,aschwanden2000b,winebarger2003}).

Several models have been proposed to explain the properties of coronal
loops at these temperatures. \cite{aschwanden2000b}, for example,
suggested that the observed loops were actually composed of smaller
scale threads that were steadily heated at their footpoints. Footpoint
heating leads to somewhat higher densities and flatter temperature
gradients relative to steady heating models. At high densities,
however, loops at these temperatures can become thermodynamically
unstable (e.g., \citealt{mok2005,muller2004,winebarger2003}), leading
to catastrophic cooling. Multi-thread, impulsive heating models have
also been suggested (e.g., \citealt{warren2003}). In these models the
fact that loops cool much more rapidly than they drain accounts for
the high densities. Multiple threads in various stages of heating and
cooling are needed to explain the observed lifetimes and temperature
gradients. In both cases, these multi-thread models indicate the need
for emission formed over a range of temperatures to reproduce the
observed intensities (see also \citealt{reale2000}).

One limitation of the observational results from \textit{TRACE} is
that they are derived from narrowband filtergrams with somewhat
limited diagnostic capabilities.  The launch of the EUV Imaging
Spectrometer (EIS) on the \textit{Hinode} mission provides us with an
opportunity to revisit some of these observational results using
spectroscopic data. EIS is a high spatial and spectral resolution
spectrometer that covers much of the same wavelength range as
\textit{TRACE}. EIS has a very broad temperature coverage and can
image the solar corona in individual emission lines from the lower
transition region to the hottest flares. 

In this paper we focus on measuring the emission measure distribution
in coronal loops near 1\,MK. We have selected 20 relatively isolated
loop segments from several different active region observations and
computed differential emission measure distributions from the
background subtracted loop intensities. For this work we focus on
loops that are bright in \ion{Fe}{12} and find that for these loops
the distribution of temperatures is almost always narrow, with a
dispersion of several times $10^5$\,K.  We also find volumetric
filling factors of approximately 10\%. These results support the idea
that coronal loops are composed of smaller scale filaments that are
below the spatial resolution of current solar instruments.

\section{Observations}

The EIS instrument on \textit{Hinode} produces high resolution
stigmatic spectra in the wavelength ranges of 171--212\,\AA\ and
245--291\,\AA. The instrument has 1\arcsec\ spatial pixels and
22.3\,m\AA\ spectral pixels. Further details are given in
\cite{culhane2007} and \cite{korendyke2006}.

From 2007 December 9 -- 18 \textit{Hinode} followed NOAA active region
10978 from near disk center to the limb. During this time EIS ran a
series of large ($460\arcsec\times384\arcsec$) slit raster
studies. The exposure time at each position in the raster was 45\,s
and each raster ran over a period of about 5 hours. The raster was
performed 9 times on this active region.

For each observation we processed the data by removing the CCD
pedestal, dark current, and hot pixels. We also estimated the
magnitude of the wavelength drift as a function of time. For each
spectral line of interest we identified line and continuum regions and
computed the line intensity, centroid, and width using
moments. Finally, we account for any spatial offsets between the two
CCDs by cross correlating rasters from emission lines formed at
similar temperatures. 

\section{Emission Measure Analysis}

\begin{figure*}[t!]
\centerline{%
 \includegraphics[clip,scale=0.76]{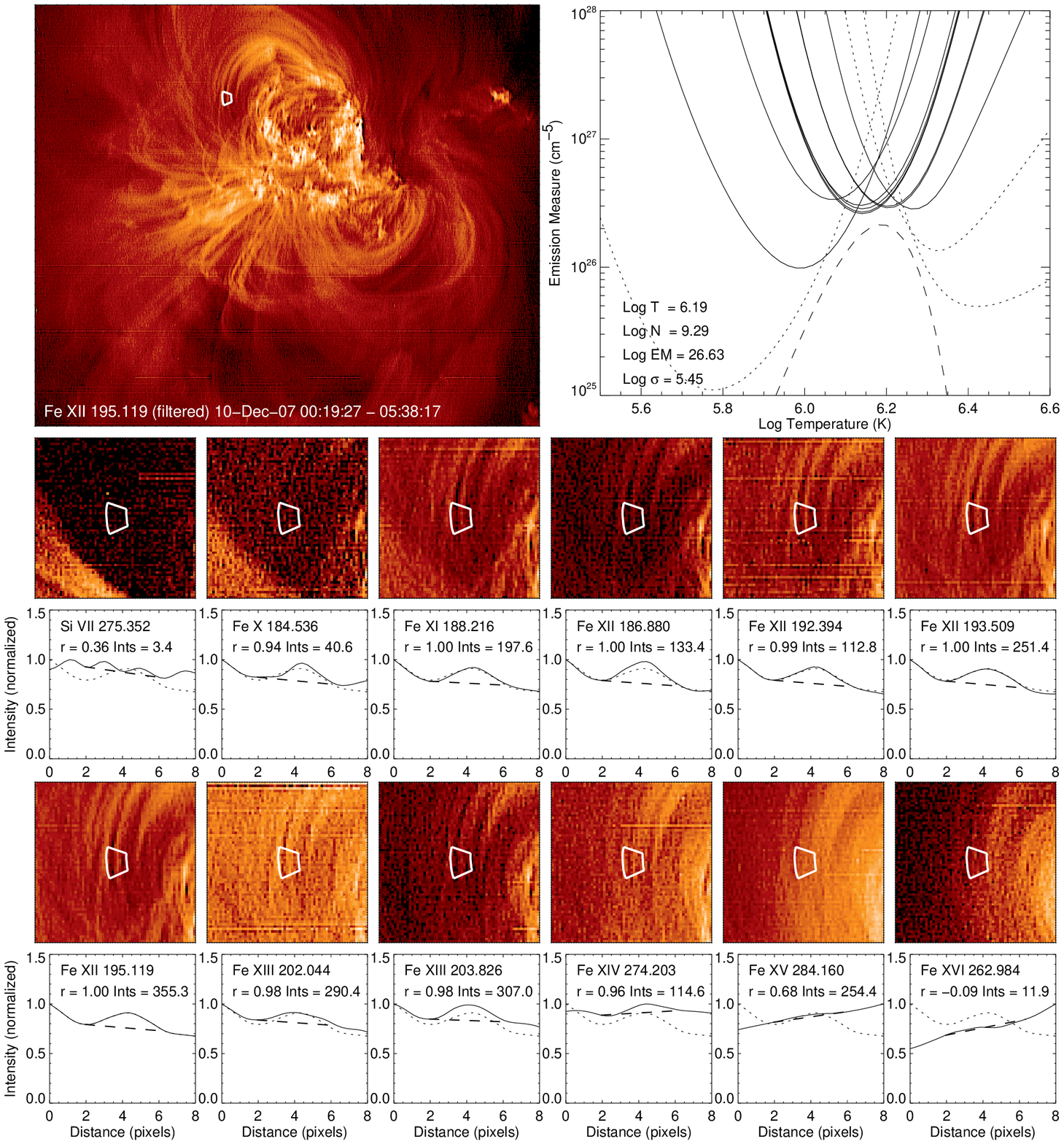}}
\caption{\small The emission measure analysis of a coronal loop segment
  observed with EIS. The region of interest is indicated with the
  white lines. The intensities averaged along the loop segment are
  also shown. For comparison the \ion{Fe}{12} 195.119\,\AA\
  intensities are repeated in each plot with the dotted line. The
  background is indicated with the dashed line. The upper right panel
  shows the EM loci for each line as well as the computed emission
  measure distribution. The dotted EM loci curves indicate that the
  intensities for these lines are not well correlated with
  \ion{Fe}{12} 195.119\,\AA. The correlation ($r$) between the
  intensity in the displayed loop intensity and the loop intensity in
  \ion{Fe}{12} 195.119\,\AA\ is given in the legend. The displayed
  images have been filtered to emphasize the contrast between the
  loops and the background emission. This is loop \# 1 in
  Table~\protect{\ref{table:em}}.}
\label{fig:loop}
\end{figure*}

\begin{figure*}[t!]
\centerline{%
 \includegraphics[clip,scale=0.76]{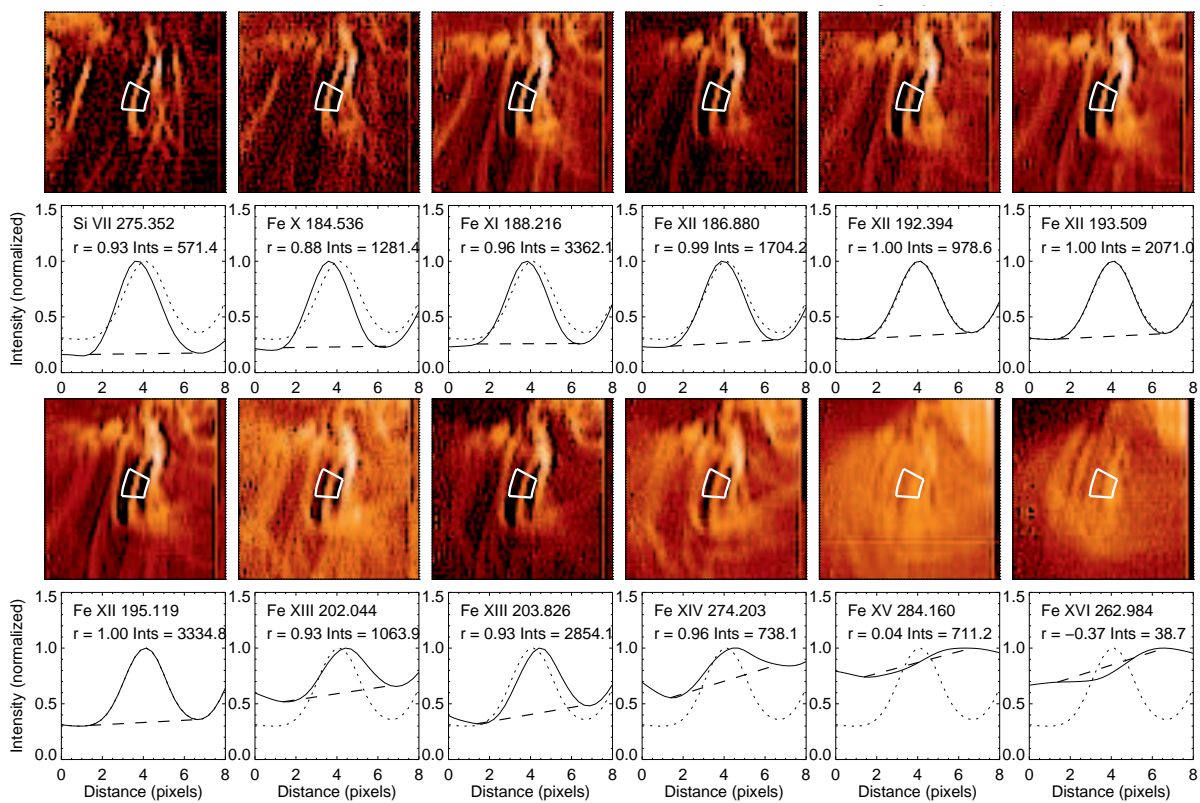}}
\caption{A loop segment observed with EIS on 2007 December 12. The
  display is similar to what is shown in
  Figure~\protect{\ref{fig:loop}}. This loop, however, appears over a
  broader range of temperatures. The field of view shown in
  $64\arcsec\times64\arcsec$. This is loop \# 5 in
  Table~\protect{\ref{table:em}}.}
\label{fig:loop2}
\end{figure*}

For this initial survey of active loops observed with EIS we inspected
each \ion{Fe}{12} 195.119\,\AA\ raster and manually identified
relatively isolated portions of coronal loops. We use the spatial
coordinates derived from this selection to determine the intensities
in the rasters of the other emission lines.  Since these loop
coordinates are not necessarily aligned to the CCD we have
interpolated to determine the intensities along the selected segment
(see \citealt{aschwanden2008b} Figure 3) and average the intensities
along the loop. Examples of EIS loop segments are shown in
Figures~\ref{fig:loop} and \ref{fig:loop2}, where the loop is shown in
various strong emission lines.

To further isolate the contribution of the loop to the observed
emission we identify background pixels in \ion{Fe}{12} 195.119\,\AA\
and fit them with a first order polynomial. The sum over the remaining
intensity between the background pixels represents the total intensity
of the loop. For consistency, these same background coordinates are
used to determine the background subtracted intensities in the other
emission lines. To determine how co-spatial the emission at the
various temperatures is, we also calculate the cross-correlation of
the background subtracted intensities with \ion{Fe}{12} 195.119\,\AA.
Note that we include 2 lines, \ion{Fe}{12} 186.880\,\AA\ and
\ion{Fe}{13} 203.826\,\AA, that form density sensitive line ratios
when paired with other lines from the same ion.

The observed background subtracted line intensities are related to the
differential emission measure in the usual way
\begin{equation}
I_\lambda = \frac{1}{4\pi}\int\epsilon_\lambda(n_e,T)\xi(T)\,dT.
\end{equation}
Since the density is an important parameter in determining the
emissivities of many of these lines, we have precomputed grids of
emissivities ($\epsilon_\lambda(n_e,T)$) as a function of temperature
and density with the CHIANTI atomic physics database (e.g.,
\citealt{landi2006}) and use the density as a free parameter in the
fitting. For the emission measure we consider two models, one a delta
function in temperature for the isothermal approximation
\begin{equation}
\xi(T) = EM_0\,\delta(T-T_0),
\end{equation}
and the other a Gaussian distribution in temperature
\begin{equation}
\xi(T) = \frac{EM_0}{\sigma_T\sqrt{2\pi}}
    \exp\left[-\frac{(T-T_0)^2}{2\sigma_T^2}\right],
\end{equation}
which allows for a dispersion in the temperature distribution. 

The calculation of the best-fit parameters for the emission measure
distributions is relatively simple. The intensities for loops that are
well correlated with \ion{Fe}{12} 195.119\,\AA\ are used directly. The
averaging generally results in very small statistical errors in the
intensities. In an attempt to account for additional uncertainties in
the atomic data we have increased the relative errors to 20\% of the
observed intensities. The intensities for emission lines that are
poorly correlated with \ion{Fe}{12} 195.119\,\AA\ ($r\le0.8$) are set
to zero. The uncertainties in these lines are estimated to be 20\% of
the measured background. The intensities and uncertainties are used as
inputs to a Levenberg-Marquardt algorithm for calculating the best-fit
parameters.

The results of applying this analysis to 20 loop segments identified
in the EIS rasters are summarized in Table~\ref{table:em}.  In almost
all cases we find that the Gaussian emission measure model has a lower
$\chi^2$ than the isothermal emission measure model. The dispersion in
temperature, however, is almost always narrow with
$\log\sigma_T\lesssim5.4$. This result is consistent with a visual
inspection of the data which shows that these \ion{Fe}{12} loops are
rarely evident in \ion{Si}{7} or \ion{Fe}{15} at the same time. In
only two cases (loops \#5 and \#11) do we see emission over such a wide
range of temperatures simultaneously.

For comparison with these active region loop measurements we have
repeated this emission measure analysis for observations above the
quiet limb, where previous work has shown the emission measure to be
isothermal (e.g., \citealt{landi2002}). In this case we obtain
$\log\sigma_T\simeq5.0$ for the Gaussian DEM model and similar
$\chi^2$ values for both the Gaussian and isothermal DEM models. The
results of this temperature analysis will be presented in a future
paper.

The assumption of a Gaussian differential emission measure is highly
restrictive. To investigate the temperature dependence of the DEM more
generally we have experimented with the Markov-chain Monte Carlo
(MCMC) reconstruction algorithm included in the PINTofALE spectroscopy
package (e.g., \citealt{kashyap2000}). The MCMC algorithm makes no
assumptions about the functional form of the DEM. The DEM computed
with this method is generally consistent with the narrow DEM suggested
by the Gaussian fits. In a number of cases the MCMC method suggests
enhanced emission above the peak temperature in the DEM. Such a
component would be consistent with the presence of cooling
filaments. However, the magnitude of the high temperature component is
sensitive to the errors assumed for the high temperature lines and
this analysis will require additional work.

For each loop segment we have also computed the loop width from a
Gaussian fit to the \ion{Fe}{12} 195.119\,\AA\ emission (see
Table~\ref{table:em}). Since we have also measured the density we can
estimate the volume of the emitting plasma in the loop. The line of
sight emission measure is simply the volume emission measure divided
by the area of an EIS pixel
\begin{equation}
EM_0 = f\frac{n_e^2V}{l^2} = f\frac{n_e^2\pi r^2l}{l^2} = 
f\frac{n_e^2\pi r^2}{l},
\label{eq:filling}
\end{equation}
where $r$ is the observed radius of the loop, $l$ is the length of EIS
pixel (1\arcsec), and $f$ is the volumetric filling factor. Following
\cite{klimchuk2000} we relate the observed loop radius to the measured
width using $r=2\sigma_w$.  The filling factors derived from the
Gaussian DEM parameters and Equation~\ref{eq:filling} are given in
Table~\ref{table:em}. This analysis suggests that these loops occupy
only about 10\% of the observed volume. 

\section{Discussion}

Some previous work has suggested that coronal loops, as currently
observed, are isothermal.  For example, \cite{aschwanden2005b} find
that the majority of narrowest loops observed with TRACE are
consistent with an isothermal DEM. Since TRACE is limited to
observations in only three channels (\ion{Fe}{9}, \ion{Fe}{12}, and
\ion{Fe}{15}) it is difficult to distinguish between an isothermal
distribution and the narrow distributions that we measure
spectroscopically. The general absence of \ion{Fe}{15} emission in the
loops that we have studied is consistent with
\cite{aschwanden2005b}. \cite{delzanna2003} also found examples of
relatively cool ($\sim0.9$\,MK) nearly isothermal loops observed with
low resolution spectroscopic data. These results also suggested
filling factors near 1. Our filling factor results are smaller than
this, but we also find that the filling factor to be inversely
proportional to the loop pressure (also see \citealt{warren2008}). We
do see some loops with a relatively broad emission measure
distribution ($\log\sigma_T\sim5.7$), which is consistent with the
results of \cite{schmelz2007} and \cite{patsourakos2007}. Our sample,
which is small, suggests that such loops are rare, however.

These new observational results lend support to the non-equilibrium,
multi-thread models of these ``warm'' coronal loops. It remains to be
seen if hydrodynamic models can reproduce the observed loop
properties. The combination of high densities and narrow temperature
ranges will be difficult to reconcile with nanoflare models (e.g.,
\citealt{patsourakos2006}). The narrow
temperature distributions suggest that these filaments are evolving
coherently.

%% ----------------------------------------------------------------------
%% --- ACKNOWLEDGMENTS --------------------------------------------------
%% ----------------------------------------------------------------------

\acknowledgments The authors would like to thank Yuan-Kuen Ko for
assistance with the MCMC DEM analysis. Hinode is a Japanese mission
developed and launched by ISAS/JAXA, with NAOJ as domestic partner and
NASA and STFC (UK) as international partners. It is operated by these
agencies in co-operation with ESA and NSC (Norway). This work was
supported by NASA and the Office of Naval Research/Naval Research
Laboratory basic research program.

%% ----------------------------------------------------------------------
%% --- REFERENCES -------------------------------------------------------
%% ----------------------------------------------------------------------

%% ----------------------------------------------------------------------
%% --- FIGURES ----------------------------------------------------------
%% ----------------------------------------------------------------------

\clearpage

\clearpage

\clearpage

\begin{deluxetable}{rrrrrcrrrcrrrrrcrrr}
\tabletypesize{\scriptsize}
\tablehead{
\multicolumn{6}{c}{} &
\multicolumn{3}{c}{Isothermal} &
\multicolumn{1}{c}{} &
\multicolumn{4}{c}{Gaussian} &
\multicolumn{4}{c}{} \\
[.3ex]\cline{7-9}\cline{11-14} \\[-1.6ex] 
\multicolumn{1}{c}{\#} &
\multicolumn{1}{c}{Date} &
\multicolumn{1}{c}{$t_{start}$} &
\multicolumn{1}{c}{$t_{end}$} &
\multicolumn{1}{c}{$\sigma_w$} &
\multicolumn{1}{c}{} &
\multicolumn{1}{c}{$EM_0$} &
\multicolumn{1}{c}{$n_0$} &
\multicolumn{1}{c}{$T_0$} &
\multicolumn{1}{c}{} &
\multicolumn{1}{c}{$EM_0$} &
\multicolumn{1}{c}{$n_0$} &
\multicolumn{1}{c}{$T_0$} &
\multicolumn{1}{c}{$\sigma_T$} &
\multicolumn{1}{c}{} &
\multicolumn{1}{c}{$\chi^2_I$} &
\multicolumn{1}{c}{$\chi^2_G$} &
\multicolumn{1}{c}{$f$(\%)}
} 
\tablewidth{0pt}
\tablecaption{Emission Measure Analysis of Active Region Loops Observed with EIS\tablenotemark{a}}
\startdata
    1 & 10-Dec-07 &  03:36:43 &  03:37:25 &    1.18  & &   26.52  &    9.25  &    6.16  & &  26.63  &    9.29  &    6.19  &    5.45  & &    1.71 &      0.79  &     9.1 \\
    2 & 11-Dec-07 &  13:11:02 &  13:11:43 &    1.42  & &   27.18  &    9.77  &    6.11  & &  27.28  &    9.86  &    6.15  &    5.44  & &    2.13 &      0.88  &     2.0 \\
    3 & 11-Dec-07 &  12:57:50 &  13:01:18 &    1.35  & &   26.90  &    9.56  &    6.13  & &  27.06  &    9.66  &    6.16  &    5.55  & &    2.86 &      1.44  &     3.3 \\
    4 & 12-Dec-07 &  06:31:29 &  06:36:21 &    1.36  & &   26.72  &    9.58  &    6.06  & &  26.79  &    9.57  &    6.07  &    5.44  & &    2.14 &      1.49  &     2.6 \\
    5 & 12-Dec-07 &  06:29:24 &  06:30:47 &    0.97  & &   27.66  &    9.61  &    6.07  & &  27.90  &    9.84  &    6.01  &    5.70  & &    5.49 &      1.52  &    19.6 \\
    6 & 12-Dec-07 &  14:52:33 &  14:53:56 &    1.17  & &   27.25  &    9.28  &    6.07  & &  27.34  &    9.43  &    6.08  &    5.54  & &    4.68 &      1.49  &    24.2 \\
    7 & 12-Dec-07 &  15:01:34 &  15:07:08 &    1.54  & &   26.62  &    9.20  &    6.08  & &  26.64  &    9.24  &    6.08  &    5.18  & &    1.42 &      1.31  &     6.8 \\
    8 & 13-Dec-07 &  15:35:17 &  15:36:41 &    1.19  & &   27.47  &    9.71  &    6.20  & &  27.49  &    9.65  &    6.20  &    5.28  & &    1.69 &      1.58  &    12.0 \\
    9 & 13-Dec-07 &  13:45:32 &  13:46:55 &    0.97  & &   26.68  &    9.34  &    6.16  & &  26.83  &    9.32  &    6.12  &    5.45  & &    3.91 &      1.65  &    18.4 \\
   10 & 15-Dec-07 &  03:40:08 &  03:41:31 &    1.03  & &   26.44  &    9.29  &    6.12  & &  26.45  &    9.31  &    6.12  &    4.99  & &    0.79 &      0.85  &     7.0 \\
   11 & 15-Dec-07 &  01:44:07 &  01:44:49 &    1.20  & &   26.64  &    9.50  &    6.13  & &  26.80  &    9.62  &    6.20  &    5.62  & &    3.73 &      3.59  &     2.8 \\
   12 & 15-Dec-07 &  21:17:07 &  21:23:22 &    2.30  & &   26.72  &    9.27  &    6.17  & &  26.77  &    9.27  &    6.16  &    5.31  & &    2.69 &      1.48  &     3.5 \\
   13 & 15-Dec-07 &  19:50:59 &  19:52:22 &    1.69  & &   26.17  &    9.39  &    6.16  & &  26.35  &    9.41  &    6.16  &    5.55  & &    1.46 &      0.85  &     1.3 \\
   14 & 18-Dec-07 &  02:15:51 &  02:17:14 &    1.07  & &   27.53  &   10.98  &    6.19  & &  27.55  &   10.50  &    6.18  &    5.44  & &    2.98 &      1.52  &     0.3 \\
   15 & 18-Dec-07 &  01:11:14 &  01:14:43 &    1.57  & &   26.51  &    9.15  &    6.19  & &  26.68  &    9.13  &    6.16  &    5.55  & &    3.16 &      1.66  &    11.5 \\
   16 & 18-Dec-07 &  01:39:43 &  01:44:35 &    2.73  & &   27.05  &    9.43  &    6.15  & &  27.14  &    9.50  &    6.17  &    5.42  & &    1.85 &      1.12  &     2.1 \\
   17 & 18-Dec-07 &  19:51:37 &  19:55:05 &    1.16  & &   26.75  &    9.86  &    6.20  & &  26.84  &    9.76  &    6.17  &    5.52  & &    1.86 &      1.34  &     1.7 \\
   18 & 10-Dec-07 &  03:27:00 &  03:32:33 &    1.28  & &   26.89  &    9.39  &    6.22  & &  26.92  &    9.34  &    6.21  &    5.36  & &    1.36 &      1.18  &    11.6 \\
   19 & 11-Dec-07 &  13:13:48 &  13:15:53 &    0.90  & &   26.60  &    9.99  &    6.19  & &  26.69  &   10.02  &    6.20  &    5.40  & &    1.00 &      0.42  &     0.6 \\
   20 & 13-Dec-07 &  16:08:38 &  16:10:01 &    1.04  & &   26.49  &    9.47  &    6.10  & &  26.58  &    9.51  &    6.09  &    5.33  & &    2.13 &      1.20  &     3.7 \\
\enddata
\tablenotetext{a}{The date and times given indicate when EIS was
  rastering over the loop segment. The paramter $\sigma_w$ is the loop
  width in pixels measured in Fe\,\textsc{xii} 195.119\,\AA. The base-10
  logarithm of the emission measure parameters are given.}
\label{table:em}
\end{deluxetable}

\end{document}